# Exploring Automated Contouring Across Institutional Boundaries: A Deep Learning Approach with Mouse Micro-CT Datasets

Lu Jiang[1], Di Xu[1], Qifan Xu[1], Arion Chatziioannou[2], Keisuke S. Iwamoto[3], Susanta Hui,[4] Ke Sheng[1, *]

1. Department of Radiation Oncology, University of California San Francisco
2. Department of Molecular and Medical Pharmacology, University of California Los Angeles
3. Department of Radiation Oncology, University of California Los Angeles
4. Department of Radiation Oncology, City of Hope

**\*** Corresponding Author: Ke Sheng

Email: ke.sheng@ucsf.edu

Address: 1600 Divisadero St., Suite H-1031, San Francisco, CA 94115

ORCID: https://orcid.org/0000-0002-6696-5409

The research is supported by NIH R01CA259008 (KS) and NIH 2R01CA154491 (SH).

**Abstract**

Image-guided mouse irradiation is essential to understand interventions involving radiation prior to human studies. Our objective is to employ Swin UNEt Transformers (Swin UNETR) to segment native micro-CT and contrast-enhanced micro-CT scans and benchmark the results against 3D no-new-Net (nnU-Net). Swin UNETR reformulates mouse organ segmentation as a sequence-to-sequence prediction task, using a hierarchical Swin Transformer encoder to extract features at 5 resolution levels, and connects to a Fully Convolutional Neural Network (FCNN)-based decoder via skip connections. The models were trained and evaluated on open datasets, with data separation based on individual mice. Further evaluation on an external mouse dataset acquired on a different micro-CT with lower kVp and higher imaging noise was also employed to assess model robustness and generalizability. Results indicate that Swin UNETR consistently outperforms nnU-Net and AIMOS in terms of average dice similarity coefficient (DSC) and Hausdorff distance (HD95p), except in two mice of intestine contouring. This superior performance is especially evident in the external dataset, confirming the model's robustness to variations in imaging conditions, including noise and quality, thereby positioning Swin UNETR as a highly generalizable and efficient tool for automated contouring in pre-clinical workflows.

1. **Introduction**

Radiotherapy (RT) treats an estimated 52.4% of all cancer patients[1] and contributes to 40% cures[2]. Despite the long history of RT, its clinical applications are still being rapidly developed with ever-improving technology, understanding of underlying biology, and emerging combined therapy[3-6]. The pre-clinical study of small animal RT response is a pivotal step to bridge the gap between in-vitro concepts and clinical practice[7,8]. With the recent advance in gene knockout and transgenic techniques, mice and other rodents have become widely used model organisms of choice in pre-clinical research given their considerable similarity shared with human physiology and pathology, versatility in genetic modification, cost-efficiency, and availability[9,10].

The past decade has continued to witness the development of RT prompted by novel small-animal irradiator-enabled comprehensive studies[11]. Small animal irradiators are commonly designed with kilovolt x-ray radiation sources combined with high-resolution 3D image guidance in the form of onboard micro-computed tomography (CT), the latter of which is used for treatment planning and target localization, parallel to modern image-guided human radiotherapy workflow[11,12]. A major progress in recent pre-clinical devices is the development of image-guided small animal irradiators, which introduces whole-body imaging modalities and advanced image-guided irradiation systems for small animals[13-15]. The Small Animal Radiation Research Platform (SARRP, Xstrahl Ltd.) and X-Rad SmART (Precision X-Ray Inc.) are two representative image-guided small animal radiation treatment systems[16,17]. Image-guided small animal irradiators empower more accurate radiation dose characterization and delivery. However, the current workflow for image-guided small animal irradiator treatment planning requires manual organ delineation, which is impractical given the limited time, resources, and expertise in the pre-clinical domain[18,19]. As a result, organ contouring is often abbreviated, inaccurately performed or even bypassed. Subsequently, the 3D dose is inadequately characterized, compromising the experimental reproducibility and translatability.

Apart from the practical challenge of manual contouring, additional challenges faced by mouse organ segmentation. First, the image-guided mouse irradiation workflow must be accomplished in real-time with the animal under anesthesia, compared to asynchronous human treatment planning. Second, a small animal irradiation experiment is often performed by a single operator whose expertise may not be mouse organ delineation, whereas human therapy is conducted by a team of experts, including those familiar with the delineation task. Third, micro-CT images for small animal irradiation are significantly noisier than high-resolution diagnostic CT for human patient treatment planning. Fourth, mouse micro-CT image quality is less uniform due to the lack of equipment and scanning protocol standards. Additionally, delineating the tumor can be particularly challenging depending on its location, especially in orthotopic grafts, making the a priori knowledge of normal organ anatomy crucial. Failing to recognize and address these challenges prevents biologists from extracting the essential dose-volume information and hampers the development of more advanced pre-clinical irradiation techniques, such as intensity-modulated radiotherapy (IMRT).

Several pioneering methods have been developed to automate small-animal organ contouring. For instance, atlas-based segmentation was used for mouse whole-body imaging[20-23]. However, its segmentation quality relies on deformable registration and the prior anatomical knowledge defined in the atlas. Following that, Van der Heyden et al.[24] developed a multi-atlas-based image segmentation (MABIS) algorithm for six organs to account for individual variations and enhance low-contrast organ segmentation. However, their proposed post-processing techniques are manual and time-demanding (~12 minutes per mouse). In addition to atlas-based approaches, Akselrod-Ballin et al.[25] proposed super-pixel machine learning algorithms learned from multiple imaging modality inputs, which can be generalized to various tissues and imaging modalities but added additional complexity to data acquisitions and required dedicated animal holders.

Deep learning (DL) has shown great promise in image processing in the past decade, including segmentation[26-28]. Multiple DL approaches were proposed for the task of mouse segmentation. Specifically, Van der Heyden et al.[29] designed a two-step 3D U-Net to automatically contour mouse skeletal muscles. Wang et al.[30] developed a 3D two-stage deeply supervised convolutional neural network (CNN) to segment multiple major organs. AIMOS[31] (AI-based Mouse Organ Segmentation) was designed to be fully automatic with several 2D U-Net-like architectures that differ in the number of encoding and decoding levels. Malimban et al.[32] applied several no-new-Net (nnU-Net) variants to segment mouse micro-CT scans and found that 3D nnU-Net models outperformed 2D models and AIMOS. Lappas et al.[33] proposed a preprocessing step that converted Hounsfield units (HUs) to mass density to improve dataset consistency, followed by a 3D U-Net for micro cone-beam mouse CT segmentation.

Fully convolutional neural networks (FCNNs), such as U-Net, have demonstrated solid performance in various medical image segmentation tasks[34-36]. However, these methods are not built on the inherent self-attention mechanism and are unstable for segmenting heterogeneous data outside the training cohort. The challenge is more significant for pre-clinical images due to the lack of standardization. Recently, Transformers, leveraging parallel learning and attention mechanisms, have demonstrated efficient and robust inferences in computer vision tasks[37-40]. Rolfe et al.[40,41] presented an open-source Mouse Embryo Multi-Organ Segmentation (MEMOS), using a fused architecture of U-Net and Transformers (UNETR). More recently, Transformers evolved to be Swin Transformers[42-45]. These models applied shifted windows self-attention scheme, referencing neighboring tokens during model propagation to enhance regional learning ability with surrounding information. This approach makes the overall network architecture more robust and generalizable. Inspired by Swin Transformers, Swin U-Net Transformers[46] (Swin UNETR) were developed specifically for generalizable medical imaging segmentation. The model outperformed other state-of-art approaches in brain tumor segmentation tasks, including nnU-Net[35], SegResNet[47] and a Vision Transformer-based model, TransBTS[48], by achieving higher Dice scores. However, the efficacy of Swin Transformer for pre-clinical micro-CT segmentation has not been studied.

Here, we introduce Swin Transformers for automatic major mouse organ contouring. Our model is trained and validated on a publicly available micro-CT dataset and compared with state-of-art models, 3D U-Net architecture, and AIMOS. A private dataset, acquired using a different micro-CT at lower kVp, was employed to assess the generalizability of our method.

## 2. Materials and Methods
### 2.1 Data
In this study, we used public and private mouse micro-CT datasets. The public dataset[49] consisted of two types of scans: native micro-CT (NACT) and contrast-enhanced micro-CT

(CECT). Specifically, the NACT dataset included 140 whole-body scans from 20 mice obtained at seven different time points using a pre-clinical micro-CT scanner (Tomoscope Duo, Germany) with an energy level of 65 kVp. The CECT dataset contained 81 scans from 8 mice, acquired at various time points with an InSyTe micro-CT scanner (BMIF TriFoil Imaging, France) at 75 kVp. Both types of public scans used an isotropic resolution of 0.14 mm. Additionally, we used a private dataset (PCT), consisting of 5 scans from 5 different mice. These images were captured using an X-RAD SmART scanner (Precision X-Ray Inc.) at 40 kVp and featured an isotropic resolution of 0.2 mm. The public and private datasets were annotated by two different biologists who were experts in delineating mouse CT anatomy. The specifics for each dataset are further detailed in Table 1.

We first employed both public NACT and CECT images to train two separate models. For the models trained on NACT, the dataset from 20 mice (140 scans) was divided into train+validation/test = 14 mice (98 scans)/6 mice (42 scans). The models trained on CECT utilized the dataset with 6 mice (66 scans) for training and validation and 2 mice (with 15 scans) for testing. This data selection was implemented to create entirely separate subsets *at the individual animal level* for training, validation, and testing, thereby ensuring an unbiased and comprehensive evaluation of the model's performance. In this study, we focused on seven organs, including the heart, lungs, liver, intestine, spleen, kidneys, and bladder.

All data were homogenized to ensure consistency as follows. Scan voxel values were normalized to [0, 1], and mouse immobilization devices were scrubbed from the background. Noteworthily, the private dataset used a lower energy acquisition setting (40 kVp) than NACT (65 kVp) and CECT (75 kVp). This resulted in a lower average signal-to-noise ratio (SNR=mean/standard deviation) for soft tissues, such as the liver, in the PCT dataset (SNR=3.57), compared to the average SNR values in the CECT (5.49) and NACT (6.59) datasets. As part of the data homogenization, the private dataset was converted to mass density using the density-HU calibration curve for 40 kVp and then converted to HU using the 65 kVp curve. The lower-resolution private dataset was linearly resampled to be 0.14 mm isotropically, the same as the public dataset.

| Dataset | Source | Scanner | Number of Animals | | Number of Images | | Energy | Resolution |
|---|---|---|---|---|---|---|---|---|
| | | | Training/ Validation | Test | Training/ Validation | Test | | |
| Native CT (NACT) | Public Dataset | Tomoscope Duo | 14 | 6 | 98 | 42 | 65 kV | 0.14 mm × 0.14 mm × 0.14 mm |
| Contrast-enhanced CT (CECT) | Public Dataset | InSyTe | 6 | 2 | 66 | 15 | 75 kV | 0.14 mm × 0.14 mm × 0.14 mm |
| Private CT (PCT) | City of Hope | SmART | / | 5 | / | 5 | 40 kV | 0.2 mm × 0.2 mm × 0.2 mm |

**Table 1.** Details of three datasets used in this study.

### 2.2 Model

#### 2.2.1 Swin Transformers for Semantic Segmentation

Swin Transformer, a variant of the general Transformers model, employs an efficient shifted window partitioning scheme, making it suited for medical image analysis where multi-scale feature extraction is important. In this study, Swin UNEt TRansformers (Swin UNETR) was adapted from Hatamizadeh et al.[46] Swin UNETR reformulates the segmentation task as a sequence-to-sequence prediction problem, where multimodal input data is projected into 1D sequences of embeddings, utilizing a hierarchical Swin Transformer as the encoder. This encoder has a patch size of 2×2×2 with 7 channels, resulting in a 56-dimensional feature space. The encoder is characterized by 4 stages with 2 transformer blocks in each, making a total of 8 layers. Swin UNETR has a U-shaped network design in which the extracted feature representations of the encoder are used in the decoder via skip connections at each resolution. At each stage of the encoder and bottleneck, the output feature representations are adjusted in size and fed into a residual block. This block consists of two 3×3×3 convolutional layers, normalized by instance normalization layers. Following this, the resolution of the feature maps is doubled using a deconvolutional layer, and the resultant outputs are concatenated with those from the preceding stage. The final segmentation is achieved using a 1×1×1 convolutional layer with a sigmoid activation function. The soft Dice loss function is applied in a voxel-wise manner. The Swin UNETR models were trained on NACT and CECT datasets separately. The inference window size is 128×128×128 with an overlap factor of 0.8 between windows. The U-shaped design incorporates Swin Transformer's strengths into a structure conducive to complex segmentation tasks, such as mouse organ segmentation for micro-CT scans. More Swin UNETR architecture details can be found in Figure 1.

### 2.2.2 nnU-Net

The nnU-Net method[35] is the first standardized out-of-the-box publicly available tool in biomedical segmentation. It is a self-adapting algorithm that selects the hyper-parameters, such as the batch size, patch size, and network topology, depending on the dataset given by the user with a set of heuristic criteria. nnU-Net offers a fully automated deep learning pipeline, including three different 3D U-Net architectures with a depth of 5. It selects the best network architecture through a 5-fold cross-validation procedure to split the data into training and validation sets. The same test set was withheld in this study, and the rest of the data were used for cross-validation. The estimated best performance of all nnU-Net models was the 3D full-resolution architecture[32]. Subsequently, the 3D full-resolution nnU-Net models were separately trained on the NACT and CECT datasets, with 32 feature channels and a batch size of 2 for comparison. More nnU-Net architecture details can be found in Figure 1.

### 2.2.3 AI-based Mouse Organ Segmentation (AIMOS)

Schoppe et al.[31] introduced AIMOS, a specialized deep-learning pipeline for segmenting mouse organs in micro-CT images. This system offers various 2D U-Net-like architectures with minimal user intervention. For this study, the UNet-768 structure was used, featuring six encoder-decoder stages with 32 and 768 feature channels at the highest layer and bottleneck, respectively. The network was trained on all slices with a batch size of 32. Malimban et al.[32] showed that 3D nnU-Net achieved better segmentation accuracy than 2D U-Net-based AIMOS for thorax organs using the same public NACT and CECT datasets. Previous studies have benchmarked nnU-Net performance against AIMOS[32]. Therefore, in this study, we used one published AIMOS model trained on the NACT dataset (*NACTmodel.pt*) to perform segmentation on the PCT dataset.

### 2.3 Evaluation metrics

The segmentation network performance was quantitatively evaluated using the Dice similarity coefficient (DSC) and the 95$^{th}$ percentile of the Hausdorff distance (HD$_{95p}$)[50] between automated and manual reference contours. The analysis is performed for individual scans and then combined for each unique animal via averaging if multiple scans are present for the same animal.

$$DSC = \frac{2|A \cap B|}{|A| + |B|} \quad (1.)$$

$$HD_{95p} = max\{d_{95}(A,B), d_{95}(B,A)\}, \quad d_{95} = x_{95} \left\{ \min_{\substack{a \in A \\ b \in B}} d(a,b) \right\} \quad (2.)$$

DSC measures the volume overlap between the references and predicted masks. A and B represent the corresponding voxels of the ground truth and the prediction, respectively. $x_{95}$ denotes the 95$^{th}$ percentile. The HD$_{95p}$ is a specific variant of the Hausdorff Distance, designed to be robust toward outliers yet relevant to radiation treatment planning, aiming at constraining most voxels to be within a certain dose level. The average DSC and HD$_{95p}$ were used for analysis to provide a balanced representation of the data across mice with varying numbers of scans, ensuring that the results were not influenced disproportionately by those with more or fewer scans.

### 2.4 Implementation details

In our project, all neural networks were trained using a single NVIDIA RTX A6000 with 48 GB of GPU memory. A five-fold cross-validation method was employed, and during each fold, two mice for NACT and one mouse for CECT were randomly selected for validation. The Adam optimizer with an initial learning rate of 0.001 was applied. The same dataset split configuration was used for all networks, and the test set and external set were withheld to evaluate and compare the predictions generated from all networks. The training process with 500 epochs took approximately 1-3 days. The inference speed was evaluated on the same system.

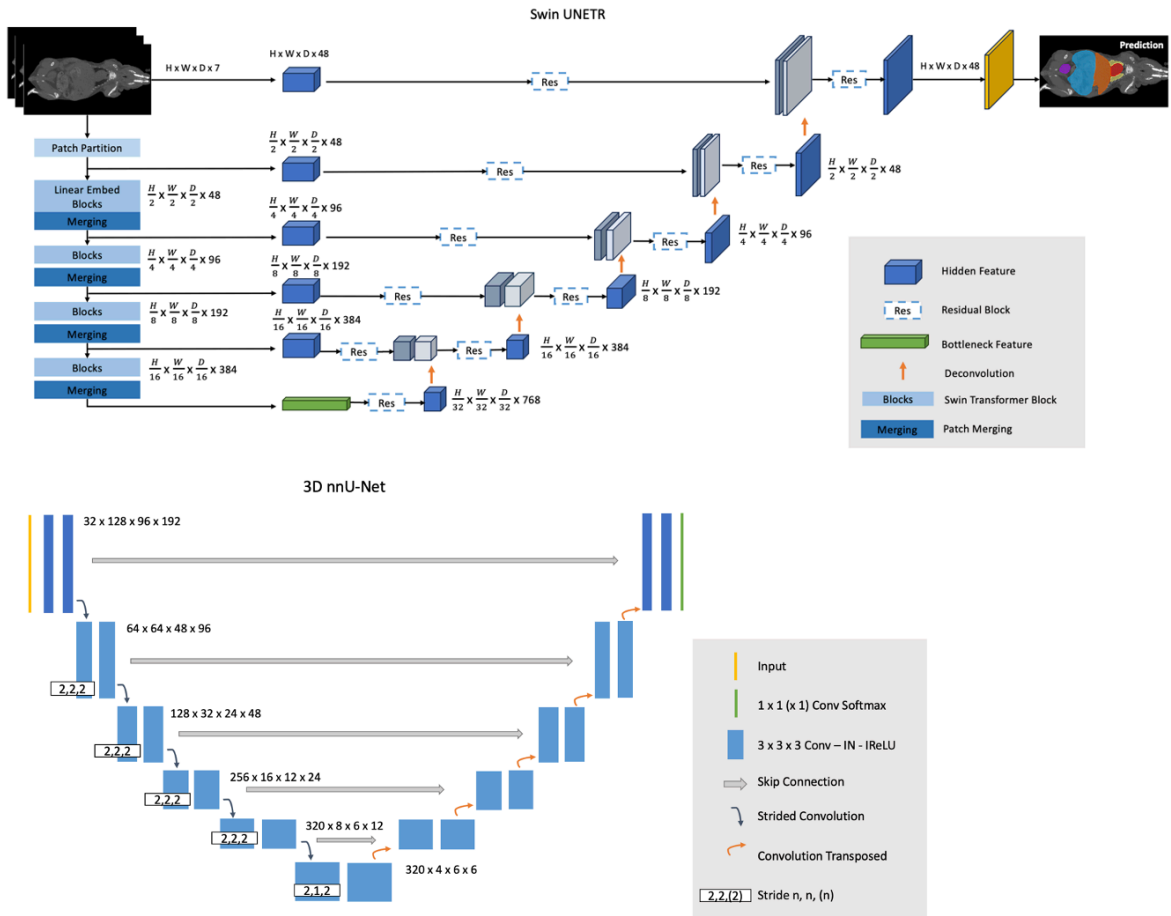

**Figure 1.** Swin UNETR architecture and 3D nnU-Net architecture used in this study.

## 3. Results

Seven major mouse organs were segmented using the Swin UNETR, 3D nnU-Net, and AIMOS. The test sets withheld from the NACT and CECT datasets were inferred and analyzed separately, and the PCT was used further to investigate each model's robustness across institutional boundaries. Figure 2 and Figure 3 illustrate comparisons for median cases between manual contouring and automated contouring by the Swin UNETR model and the 3D nnU-Net model from the NACT and CECT test sets. Generally, both neural networks accurately segmented the target organs, except for the intestine and spleen. The spleen lacked contrast in the NACT dataset but was visible in the CECT dataset.

Additionally, the automatically delineated boundaries of predictions from both models were smoother compared to the actual ground truth, a characteristic most evident in the lungs. Figure 4 shows the model's generalizability to the PCT using completely different imaging equipment, protocol, and organ annotator. Swin UNETR was better at capturing lung features and providing more precise boundary predictions for the bladder, liver, and kidneys than the other two models.

Both average DSC and $HD_{95p}$ were reported on the individual animal level for the NACT and CECT test sets in Table 2. For the NACT test set, Swin UNETR generally showed slightly higher DSC in most organs, except in the intestine, where nnU-Net performed marginally better in two mice. Consistently, both neural networks had difficulties with spleen segmentation in the NACT dataset, resulting in approximately 70% DSC and ~1 mm $HD_{95p}$. However, for the CECT test set, contrast agents significantly enhanced spleen segmentation, achieving more than 90% DSC and ~0.6mm $HD_{95p}$. All models achieved $HD_{95p}$ less than 1 mm except for the liver and intestine in both NACT and CECT test sets and the spleen in the NACT test set. For the PCT dataset, metrics were reported for 5 mice in Table 3. Superior performance and generalization using Swin UNETR, compared to the other two models trained with the NACT dataset, were more pronounced in the completely unseen PCT dataset. Swin UNETR consistently achieved superior DSC and $HD_{95p}$ for all 7 organs, providing more than 80% DSC in the bladder, lungs, and liver, while the other two models' performance suffered an evident drop to ~70% DSC. For kidneys, the DSC also improved from ~60% for the 3D nnU-Net and AIMOS, to 73.1% for Swin UNETR. The $HD_{95p}$ was nearly halved to 1.44 mm vs. 2.81mm for AIMOS.

The yellow arrows in Figures 2-4 denote key differences in segmentation performance across various models. Specifically, arrows in Figures 2-4 indicate that all deep-learning models yield smoother boundary contours for the lungs and mis-segmentations in the intestine. In Figure 2 for the NACT dataset, arrows show over-segmentations at the spleen boundary due to a lack of contrast with adjacent tissues. In Figure 4 for the PCT dataset, more false-positive islands can be observed in heart, liver, and kidney segmentation, as well as minor false-negative islands in the bladder.

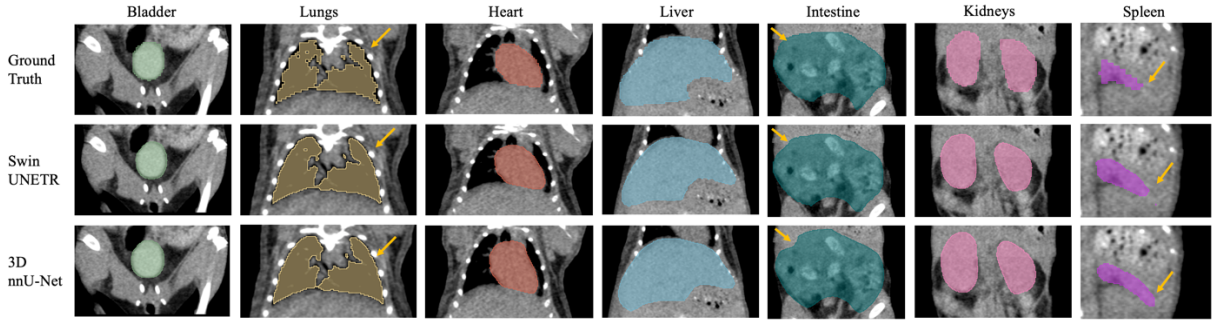

**Figure 2.** Example of the median-scored case in mouse multi-organ segmentation in coronal view from the NACT test set. Yellow arrows denote key differences in segmentation.

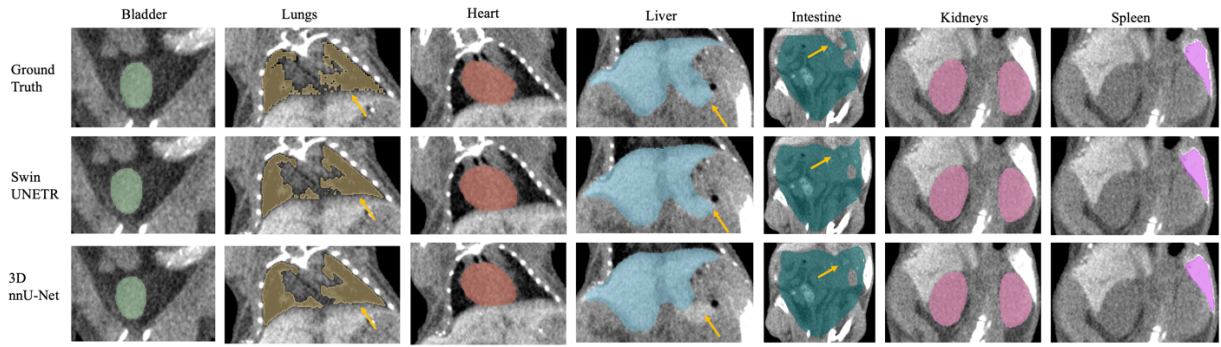

**Figure 3.** Example of the median-scored case in mouse multi-organ segmentation in coronal view from the CECT test set. Yellow arrows denote key differences in segmentation.

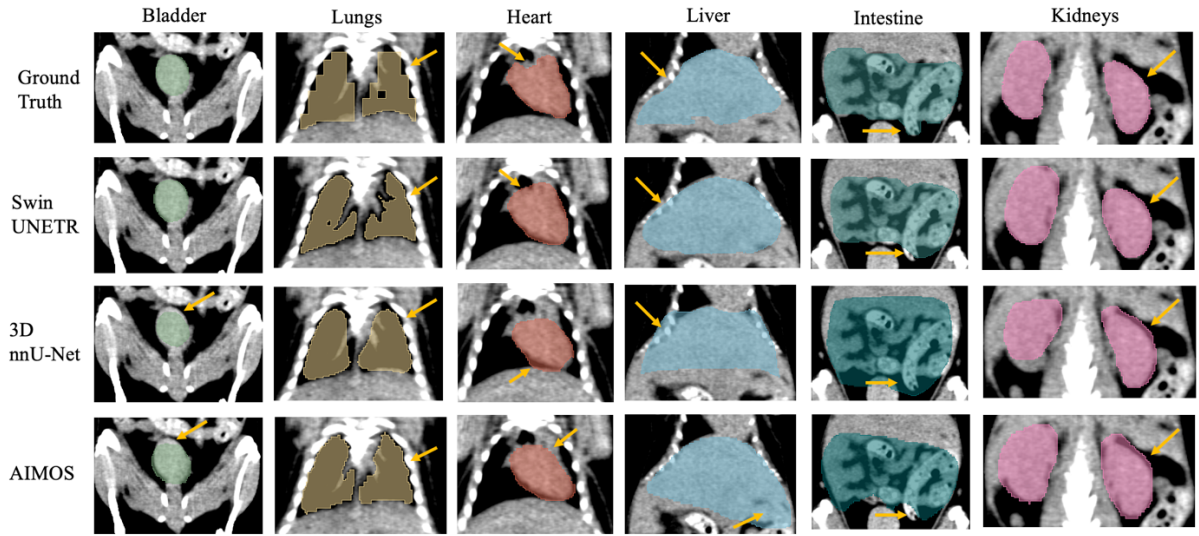

**Figure 4.** Example of the median-scored case in mouse multi-organ segmentation in coronal view from the PCT test set. Yellow arrows denote key differences in segmentation.

| DSC (%) mean±s.d. | | Bladder-1 | | Lungs-2 | | Heart-3 | | Liver-4 | | Intestine-5 | | Kidneys-6 | | Spleen-7 | |
|---|---|---|---|---|---|---|---|---|---|---|---|---|---|---|---|
| | | *Swin UNETR* | *nnU-Net* | *Swin UNETR* | *nnU-Net* | *Swin UNETR* | *nnU-Net* | *Swin UNETR* | *nnU-Net* | *Swin UNETR* | *nnU-Net* | *Swin UNETR* | *nnU-Net* | *Swin UNETR* | *nnU-Net* |
| NACT | M01 | **90.9±1.8** | 90.1±2.3 | **89.4±1.5** | 88.3±1.3 | **92.6±1.3** | 90.8±1.2 | **91.1±1.2** | 88.8±1.3 | **86.5±2** | 86±1.8 | **91±1.1** | 90.4±1 | **74.6±7.7** | 72.5±8.7 |
| | M02 | **92.5±2.3** | 92.4±2 | **91.2±0.9** | 89.2±0.9 | **91.8±1.2** | 90.8±1.3 | **92±1.3** | 90.5±1.1 | **89.7±1.9** | 89.3±1.8 | **91.3±1.4** | 90.4±1.3 | **75.2±3.6** | 73.8±6.5 |
| | M03 | **92.3±0.6** | 90.7±1 | **90.6±1.2** | 89.4±1.5 | **91.4±0.9** | 91.3±1.1 | **91±0.5** | 90.9±1.1 | 88.7±1.5 | <u>**88.9±1.2**</u> | **88.8±2** | 88.6±3.1 | **75.5±3.9** | 74.9±4.2 |
| | M04 | **91.7±1.6** | 90.5±1.5 | **92.5±0.7** | 89.8±0.9 | **92.7±0.6** | 91.6±0.5 | **91.6±1** | 91.1±1.2 | 89.8±1.3 | 89.8±1.5 | **90.9±1.1** | 90.6±1 | **76.1±4.4** | 74.1±6 |
| | M05 | **88.5±0.6** | 86.8±1.3 | **89.5±0.7** | 88.9±0.9 | **93.4±0.3** | 92.1±0.4 | **92±1.3** | 90.2±1.4 | 88.2±1.6 | <u>**88.4±1.7**</u> | **90±1.4** | 89.9±1.7 | **77.2±2.8** | 75.8±4.5 |
| | M06 | **89.8±2.6** | 89.1±2.3 | **93.2±1.6** | 92.4±1.5 | **93±0.6** | 91.8±0.8 | **87.7±0.7** | 87.6±1.8 | 88±1.5 | 88±2.2 | **86.6±2.7** | 85.5±3 | **67.5±7.2** | 65.9±8.7 |
| CECT | M01 | **91.9±3** | 90.1±3.2 | 89.9±4 | 89.9±4.8 | **93±1.3** | 92.3±1.6 | **92.7±0.3** | 92.4±0.6 | **89.5±2.9** | 88.9±2.8 | **92±1.3** | 91.2±1.3 | **92.4±1.4** | 90.5±2.5 |
| | M02 | **91.4±3.1** | 89.1±3 | **86.6±5.5** | 85.8±7.1 | **92.5±1.4** | 90.8±1.6 | **88.5±4.2** | 85.5±4.3 | **84.9±3.5** | 84.6±5.1 | **88.8±2.8** | 87.7±2.6 | **92.3±1.3** | 91.4±1 |

| HD$_{95p}$ (mm) mean±s.d. | | Bladder-1 | | Lungs-2 | | Heart-3 | | Liver-4 | | Intestine-5 | | Kidneys-6 | | Spleen-7 | |
|---|---|---|---|---|---|---|---|---|---|---|---|---|---|---|---|
| | | *Swin UNETR* | *nnU-Net* | *Swin UNETR* | *nnU-Net* | *Swin UNETR* | *nnU-Net* | *Swin UNETR* | *nnU-Net* | *Swin UNETR* | *nnU-Net* | *Swin UNETR* | *nnU-Net* | *Swin UNETR* | *nnU-Net* |
| NACT | M01 | **0.42±0.16** | 0.47±0.16 | **0.29±0.11** | 0.31±0.16 | **0.42±0.2** | 0.46±0.24 | **1.16±0.47** | 1.38±0.33 | 1.86±0.66 | 1.86±0.53 | **0.57±0.2** | 0.65±0.11 | **0.97±0.32** | 1.11±0.44 |
| | M02 | **0.39±0.17** | 0.41±0.17 | 0.28±0.09 | 0.28±0.08 | 0.47±0.14 | 0.47±0.16 | **0.93±0.37** | 1.03±0.39 | **1.86±0.46** | 1.93±0.68 | **0.56±0.18** | 0.69±0.12 | **0.85±0.19** | 1.15±0.29 |
| | M03 | **0.41±0.14** | 0.45±0.13 | **0.27±0.07** | 0.48±0.11 | **0.37±0.15** | 0.40±0.16 | **0.98±0.28** | 1.03±0.62 | 2.31±0.68 | <u>**2.28±0.7**</u> | **0.61±0.29** | 0.75±0.24 | **1.06±0.19** | 1.03±0.17 |
| | M04 | **0.43±0.12** | 0.46±0.13 | **0.27±0.12** | 0.30±0.14 | 0.39±0.1 | 0.39±0.17 | **0.99±0.3** | 1.02±0.38 | **1.60±0.4** | 1.63±0.58 | **0.53±0.14** | 0.65±0.15 | **1.00±0.23** | 1.09±0.34 |
| | M05 | 0.64±0.17 | 0.64±0.19 | **0.26±0.1** | 0.31±0.15 | **0.36±0.07** | 0.37±0.08 | **0.98±0.29** | 1.08±0.3 | 1.98±0.33 | <u>**1.91±0.45**</u> | **0.58±0.18** | 0.68±0.17 | **1.06±0.21** | 1.12±0.32 |
| | M06 | **0.37±0.14** | 0.51±0.16 | 0.32±0.18 | 0.32±0.2 | 0.39±0.16 | 0.39±0.2 | **1.23±0.53** | 1.47±0.66 | **2.02±0.51** | 2.03±0.62 | **0.65±0.15** | 0.77±0.25 | **1.39±0.3** | 1.65±0.45 |
| CECT | M01 | **0.38±0.1** | 0.44±0.2 | **0.25±0.15** | 0.29±0.13 | **0.52±0.24** | 0.58±0.19 | **1.14±0.29** | 1.26±0.42 | **1.78±0.5** | 1.94±0.67 | **0.62±0.29** | 0.65±0.24 | **0.72±0.2** | 0.76±0.33 |
| | M02 | **0.46±0.11** | 0.50±0.14 | **0.27±0.13** | 0.35±0.1 | **0.59±0.27** | 0.62±0.2 | **1.06±0.39** | 1.10±0.5 | **2.04±0.9** | 2.10±0.8 | **0.59±0.2** | 0.83±0.38 | **0.64±0.12** | 0.73±0.26 |

**Table 2.** Quantitative evaluation results on average DSC (%), HD$_{95p}$ (mm) for the NACT and CECT test sets. Swin UNETR model performance is compared with the 3D full-resolution nnU-Net model. The better results are bolded, with those not from Swin UNETR also underlined.

|  | Bladder-1 | | | Lungs-2 | | | Heart-3 | | | Liver-4 | | | Intestine-5 | | | Kidneys-6 | | |
|---|---|---|---|---|---|---|---|---|---|---|---|---|---|---|---|---|---|---|
| mean±s.d. | Swin UNETR | nnU-Net | AIMOS | Swin UNETR | nnU-Net | AIMOS | Swin UNETR | nnU-Net | AIMOS | Swin UNETR | nnU-Net | AIMOS | Swin UNETR | nnU-Net | AIMOS | Swin UNETR | nnU-Net | AIMOS |
| DSC (%) | **86.9±4.1** | 74.2±5.4 | 72.6±11 | **84.8±2.5** | 78±2.1 | 72.4±2.1 | **89.9±1.2** | 86±3.1 | 86.3±1.8 | **84.1±1.4** | 75.2±2.6 | 68.9±2.5 | **76.9±2.2** | 72±3.9 | 63.7±4.8 | **73.1±6.9** | 64.2±9.2 | 58.9±8 |
| $HD_{95p}$ (mm) | **0.62±0.17** | 0.89±0.58 | 2.1±1.14 | **0.69±0.13** | 1.22±0.14 | 0.94±0.06 | **0.6±0.06** | 1±0.14 | 0.79±0.08 | **1.8±0.18** | 1.97±0.66 | 2.24±1.1 | **2.57±0.7** | 2.88±0.37 | 3.86±1.25 | **1.44±0.5** | 2.79±1.2 | 2.81±0.41 |

**Table 3.** Quantitative evaluation results on average DSC (%), $HD_{95p}$ (mm) for the ECT test set. Swin UNETR model performance is compared with the 3D full-resolution nnU-Net and the published AIMOS model. The better results are bolded, with those not from Swin UNETR also underlined.

## 4. Discussion

3D image-guided pre-clinical irradiation platforms afford more accurate and conformal dose delivery for targeted interventional response assessment. The conformal dose delivery also lends to more translatable radiation research[51]. However, conformal dose distribution needs to be contextualized with 3D organ contours, which are not readily available in the existing pre-clinical research workflow. The importance of accurate organ and structure contours increases with the recently introduced small animal intensity-modulated radiotherapy (IMRT), which better mimics human radiotherapy[52-57]. IMRT planning solves an inverse optimization problem for specific 3D organ dosimetric goals, thus requiring accurate delineation of involved organs. Manual delineation of normal organs has been routinely performed for human patients, but such a task can be impractical for pre-clinical research. Automated segmentation of the mouse organs has been performed using conventional methods, including active contouring and deformable registration. DL has emerged as a more precise tool for organ segmentation in mouse imaging. Several existing studies have employed CNNs, e.g., 2D and 3D U-Net, to facilitate automated segmentation of mouse organs. Despite their improved performance over conventional methods, CNNs lack inherent self-attention mechanisms necessary for stable performance, as shown in "domain shift" problems[58]. Swin Transformers demonstrated superior robustness with the shifted window self-attention mechanism and a hierarchical architecture for natural image processing. In this study, we focused on investigating the robustness of Swin Transformers versus U-Net variants (nnU-Net and AIMOS)[31,35].

Swin UNETR consistently outperformed the 3D nnU-Net model except for two mice, where the latter performed marginally better on the intestine, which has an intrinsically large manual delineation uncertainty due to morphologically complex and heterogeneous imaging intensity [31,32]. This uncertainty was observed from inter- and intra-observer variations in the ground truth contours across the three test sets. Additionally, the intestine contouring can be highly subjective, lacking clear boundaries to surrounding tissues, especially in low-resolution and noisy CBCT images. Spleen segmentation posed a challenge for both neural networks in the NACT dataset, for the spleen often exhibits low contrast relative to adjacent tissues, such as the stomach, pancreas, and kidneys. Notably, applying contrast agents in the CECT test set yielded significant improvements in spleen segmentation, achieving over 90% DSC as illustrated in Figure 3, where the spleen appears highlighted. Although contrast agents did influence the liver, there were no discernible improvements in the CECT dataset. In NACT images, the liver typically exhibits well-defined boundaries and relatively high contrast with adjacent tissues, which allows for accurate segmentation by neural networks. These intrinsic characteristics of the liver may be sufficient for precise segmentation, rendering the additional contrast agents used in CECT less impactful.

Domain shift is one of the crucial tests to measure DL model generalizability[58], a property essential for the pre-clinical micro-CT images demonstrating substantial inter-institution variation due to the lack of standards. Imaging equipment, scanning geometry, kVp, and mAs can substantially alter the CT image characteristics. Yet, existing models for mouse segmentation are often trained and tested on samples acquired within one institution with homogeneous scanning parameters. The untested model robustness can hinder the adoption of automated segmentation. We acquired a PCT dataset using different micro-CT and image protocols to test the model robustness, including a lower kVp and resultant lower SNR. A different expert annotated the images. Inter-observer variations led to an estimated average decline in DSC of 8% and $HD_{95p}$ of 0.5 mm for Swin UNETR, compared to a 15% and 1 mm decline for 3D nnU-Net. Both models were trained on the NACT dataset with the original annotator. The declines in DSC and $HD_{95p}$ were calculated by comparing the models' average performance on the PCT dataset, annotated by a different individual, against their performance on the NACT dataset. This comparison provided an approximate measure of annotator bias, showing the performance drop when models trained on one annotator's labels were evaluated against another. Swin UNETR demonstrated its robustness to domain shift and consistently outperformed 3D full-resolution nnU-Net and AIMOS. The superior performance of Swin UNETR can be firstly attributed to its ability to capture long-range dependencies through an inherent self-attention mechanism, allowing for accurate organ structure recovery. Second, the model effectively balances global context awareness with local feature extraction by integrating Swin Transformers with a U-shaped network, further facilitating overall and detailed contour representation understanding.

This study is not without limitations. First, the public dataset used in this study included 221 sequential images of only 28 mice. Although we reasonably assumed the anatomical similarity of these genetically homogeneous mice, the training was not based on fully independent samples. To mitigate the limitation, we carefully split data based on individual mice to minimize interdependence among the training samples. In the current work on the other hand, the concern of data dependence is largely answered by the independent test on PCT. Second, all NACT, CECT and PCT datasets scans were conducted in the prone position. Mice setup in other postures, including supine and decubitus positions, likely require training a new model on corresponding CT images. Third, neural network performance is inherently task-specific and contingent upon the data used, as different DL methods yield varying results on disparate datasets. Fourth, auto-segmentation on low native contrast organs (spleen) and morphologically complex organs (intestine) can be unreliable. Possible solutions, including domain adaptation adversarial generative networks[59], could be explored to highlight spleen-related voxels to assist neural networks further. Lastly, the test results depend on the quality and consistency of manual annotation. Digitization errors, such as the rough boundaries of the manual lung contours, contributed to residual discrepancies that cannot be completely eliminated.

## 5. Conclusion

In this study, we assessed the performance of Swin UNETR in segmenting major mouse organs across different datasets. Swin UNETR consistently outperformed 2D and 3D U-Net. Most importantly, Swin UNETR demonstrated superior robustness via testing on an independent mouse CT dataset with substantially different image characteristics. The resilience to more noisy images is an important step toward a generalizable auto-segmentation method for pre-clinical radiation research.

**Funding**

The research is supported by NIH R01CA259008 (KS) and NIH 2R01CA154491 (SH).

**Data Availability Statement**

All data and models are available in the paper.

**Author contributions**

L.J., D.X., Q.X. designed the model and computational framework. L.J. carried out the implementation. S. H., A.C. and K.I. reviewed methods and results. K.S. supervised the project. All authors contributed to the manuscript writing.

**Conflicts of Interest**

The authors declare no conflict of interest.